\documentclass[10pt,twocolumn,letterpaper]{article}

\usepackage{iccv}              
\usepackage{multirow}

\definecolor{iccvblue}{rgb}{0.21,0.49,0.74}
\usepackage[pagebackref,breaklinks,colorlinks,allcolors=iccvblue]{hyperref}
\usepackage[accsupp]{axessibility}  

\def\paperID{} 
\def\confName{ICCV}
\def\confYear{2025}

\title{TAGS: 3D Tumor-Adaptive Guidance for SAM}

\author{
Sirui Li$^{1}$\quad 
Linkai Peng$^{2}$ \quad 
Zheyuan Zhang$^{2}$ \quad 
Gorkem Durak$^{2}$ \quad 
Ulas Bagci$^{2}$ \\ 
\small $^{1}$Southern University of Science and Technology \quad $^{2}$Northwestern University \\
\tt\small 12012719@mail.sustech.edu.cn, \\
\tt\small \{linkai.peng, zheyuan.zhang, gorkem.durak, ulas.bagci\}@northwestern.edu
}

\begin{document}
\maketitle
\begin{abstract}
Foundation models (FMs) such as CLIP and SAM have recently shown great promise in image segmentation tasks, yet their adaptation to 3D medical imaging—particularly for pathology detection/segmentation—remains underexplored. A critical challenge arises from the domain gap between natural images and medical volumes: existing FMs, pre-trained on 2D data, struggle to capture 3D anatomical context, limiting their utility in clinical applications like tumor segmentation. To address this, we propose an adaptation framework called \textbf{TAGS}: \textbf{T}umor \textbf{A}daptive \textbf{G}uidance for \textbf{S}AM, which unlocks 2D FMs for 3D medical tasks through multi-prompt fusion. By preserving most of the pre-trained weights, our approach enhances SAM’s spatial feature extraction using CLIP’s semantic insights and anatomy-specific prompts. Extensive experiments on three open-source tumor segmentation datasets prove that our model surpasses the state-of-the-art medical image segmentation models (+46.88\% over nnUNet), interactive segmentation frameworks, and other established medical FMs, including SAM-Med2D, SAM-Med3D, SegVol, Universal, 3D-Adapter, and SAM-B (at least +13\% over them). This highlights the robustness and adaptability of our proposed framework across diverse medical segmentation tasks. Our code and model are available at: \url{https://github.com/sirileeee/TAGS}.
\end{abstract}
    
\section{Introduction}
\label{sec:intro}

\begin{figure}[htbp]
\centering
\setlength{\belowcaptionskip}{-0.6cm}   
\centerline{\includegraphics[height=5.9cm]{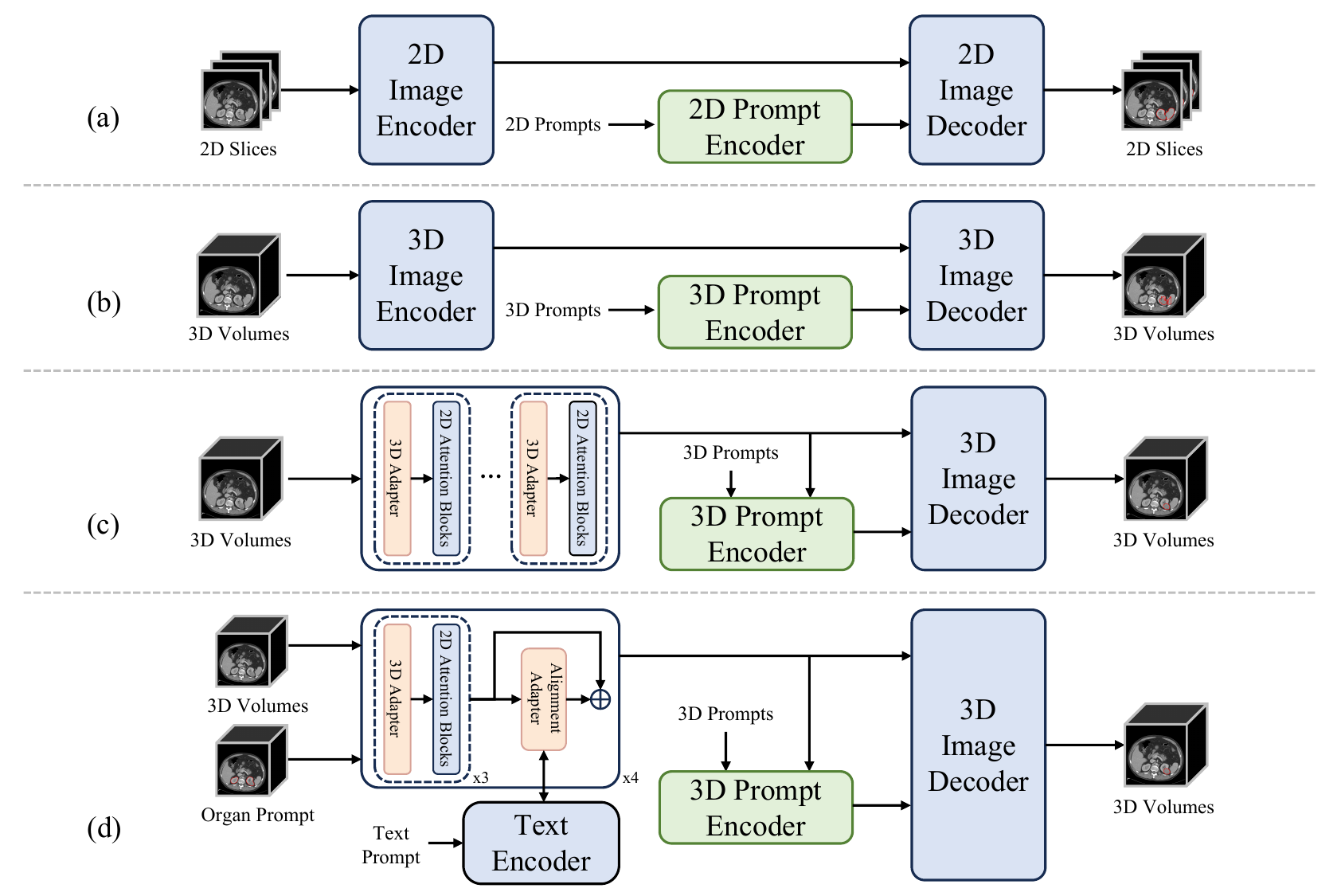}}
\caption{Comparison of 4 approaches for SAM-driven volumetric tumor segmentation: (a) slice-by-slice segmentation using 2D models; (b) employing 3D medical foundation models; (c) adapting 2D models for 3D segmentation; (d) our proposed TAGS, featuring tumor-specific adaptations.}
\medskip
\vspace{-0.2cm}
\label{fig1}
\end{figure}

Medical image segmentation is essential for applying advanced Artificial Intelligence (AI) technologies in clinical settings, increasing diagnostic accuracy, and optimizing personalized treatment planning \cite{medicalseg-important1, medicalseg-important2}. Tumor segmentation, in particular, remains one of the most challenging tasks due to the ambiguous boundaries and varying sizes and locations of tumors, along with other unique challenges related to scanners and patient population differences \cite{firstintro1, firstintro2}.

While convolutional neural networks (CNNs), particularly U-Net \cite{unet} and its variants \cite{unet-variant1, unet-variant2}, have dominated medical segmentation by leveraging localized feature extraction, their limited capacity to model global contextual relationships often leads to suboptimal delineation of tumors from surrounding tissues. In contrast, Transformer-based architectures \cite{transformer} address this issue through self-attention mechanisms to model long-range dependencies, effectively capturing global context correlations. Thus, they have demonstrated their strengths in tumor segmentation \cite{transformer-tumorseg1, transformer-tumorseg2, transformer-tumorseg3}, while still with weaknesses: their computational complexity and reliance on large-scale annotated datasets hinder practical deployment in clinical workflows. 

Recently, transformer-based foundation models \cite{fm1, fm2, fm3} have made substantial breakthroughs in computer vision, particularly in segmentation tasks through promptable inference and generalization to unseen data. With large-scale pre-training, these models perform better across diverse downstream tasks and handle unseen data more efficiently. The Segment Anything Model (SAM)~\cite{sam} exemplifies this trend with its interactive object-prompting mechanism. However, studies have shown that applying SAM directly to medical image segmentation presents unique challenges \cite{samseg1, samseg2}. First, medical images are typically 3D volumetric data, while SAM is pre-trained on 2D images, limiting its capacity to capture 3D spatial information effectively. Although SAM2 \cite{sam2}, designed for video segmentation, provides additional contextual dimensions, it remains insufficient for medical segmentation tasks \cite{sam2cannotmedseg}. Furthermore, the significant domain gap between natural and medical images constrains SAM’s ability to generalize effectively to medical data.

However, applying SAM directly to 3D medical imaging—particularly tumor segmentation—faces two critical limitations~\cite{samseg1, samseg2}. First, SAM’s 2D pre-training on natural images fails to capture the inter-slice spatial relationships inherent to volumetric medical data, resulting in inconsistent predictions across adjacent slices. Although SAM2~\cite{sam2}, designed for video segmentation, provides an additional contextual (temporal) dimension, it remains insufficient for medical segmentation tasks~\cite{sam2cannotmedseg}. Second, the semantic-contextual misalignment between natural and medical domains—such as the hierarchical dependency between organs and tumors—undermines SAM’s ability to localize pathology accurately. 


Prior methods have explored several approaches, as illustrated in~\cref{fig1}. One approach is slice-by-slice segmentation, which decouples volumetric images into 2D slices~\cite{slice1, slice2, slice3}. It applies SAM to each slice individually and then stacks the segmentation results to form a 3D output. This method, however, requires extensive prompting for each slice and fails to capture inter-slice correlations, which leads to inconsistent predictions across the volume. Another approach aims to develop 3D medical segmentation foundation models~\cite{3DSAM1, 3DSAM2}. Nonetheless, the scarcity of publicly accessible 3D medical datasets with tumor annotations limits these models. As a result, they are often pre-trained for both organ and tumor segmentation together. Yet, the difficulty in aligning data from diverse sources and the high computational demands for pre-training further limit this approach's feasibility.

Incorporating 3D adapters into the original SAM architecture has also proven effective in this manner~\cite{SAM-Adapter1, SAM-Adapter2, SAM-Adapter3}. By adding 3D adapters, the FM captures spatial information while retaining the majority of SAM's pre-trained weights. Throughout the process, SAM’s 2D encoders stay frozen, which minimizes the number of trainable parameters. This setup accelerates both model development and deployment, enhancing application speed. However, the lack of spatial awareness in the original pre-trained SAM can limit its ability to fully capture volumetric features.

To address all these challenges, we propose TAGS, a 3D \textbf{T}umor-\textbf{A}daptive \textbf{G}uidance for \textbf{S}AM, a parameter-efficient framework that adapts SAM’s 2D capabilities to 3D tumor segmentation through synergistic integration of CLIP’s semantic priors \cite{clip} and anatomy-specific prompting. Unlike prior works that naively fuse CLIP’s feature maps with SAM \cite{clipsam1, clipsam2}, TAGS employs a hierarchical prompting mechanism where CLIP’s text-guided embeddings refine SAM’s spatial attention to prioritize anatomically relevant regions. Pixel-wise loss tackles the foreground-background pixel imbalance in volumetric tumor segmentation, optimizing learning efficiency. Our organ-specific prompts, generated via automated segmentation tools, constrain SAM’s focus to tumor-prone anatomical contexts, mitigating false positives caused by spatial uncertainty. 


Our major contributions are threefold:
\begin{itemize}
    \item We propose the first multi-level feature alignment framework to unify SAM’s spatial acuity with CLIP’s semantic embeddings for 3D tumor segmentation, requiring only 18\% of SAM-B’s tunable parameters. This hierarchical prompting mechanism bridges the 2D-to-3D domain gap while preserving computational efficiency.
    \item To overcome the spatial uncertainty of tumors, we design an organ-specific prompt module that dynamically localizes tumors within anatomically plausible regions, directing the model to segment tumors within specific organ regions and reduce false positives.
    \item Extensive experiments on multiple datasets (Liver, Kidney, and Pancreas Tumors) consistently validate our approach, achieving new state-of-the-art results in volumetric tumor segmentation, with an average Dice score improvement of +46.88\% over nnUNet \cite{nnunet} and at least +13\% over medical FMs \cite{3DSAM1, 3DSAM2}.
\end{itemize}
\section{Related Work}
\label{sec:relatedwork}

\begin{figure*}[htbp]
\centering
\vspace{-0.9cm}
\setlength{\abovecaptionskip}{-0.1cm}   
\setlength{\belowcaptionskip}{-0.2cm}   
\centerline{\includegraphics[height=7 cm]{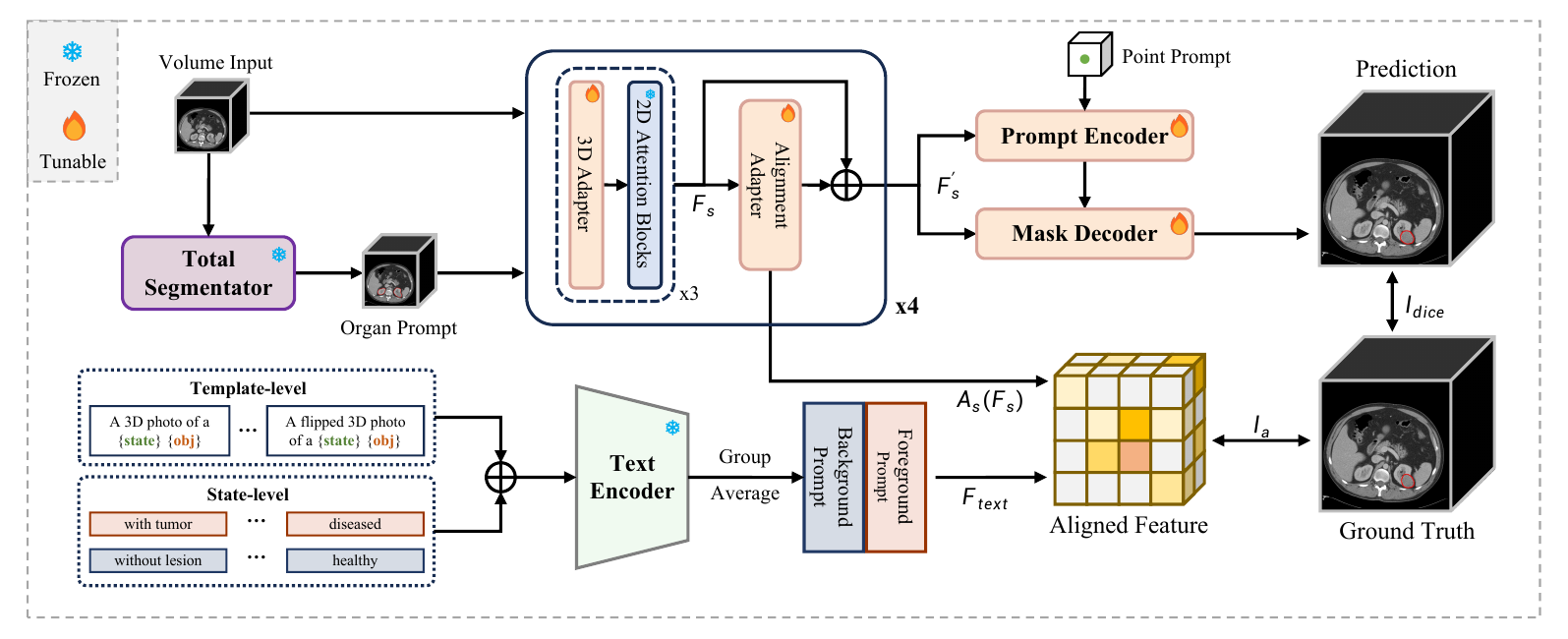}}
\caption{The overall framework of proposed model (TAGS). TAGS includes 3 different prompts: (a) annotation-free organ prompt; (b) dual-category text prompt generated by CLIP text encoder; (c) interactive 3D point prompts. Organ prompt is input into the image encoder with the raw image. Stage-wise alignment adapters then integrate foreground and background semantic information from the text prompt into the image feature extraction. Finally, the image features merge with the point prompt, and the mask decoder predicts the tumor region.}
\vspace{-0.1cm}
\label{fig2}
\end{figure*}

\subsection{Volumetric Tumor Segmentation}
Volumetric tumor segmentation is fundamental for oncology workflows, including accurate cancer diagnosis, staging, and prognosis. In comparison to 2D segmentation, volumetric (3D) approaches must overcome increased computational overhead, higher data dimensionality, and the complexity of delineating ambiguous tumor boundaries across multiple slices. Classic 3D methods such as 3D U-Net \cite{3dunet}, TransUnet \cite{transunet}, UNetr \cite{unetr}, and Cotr \cite{cotr} have set benchmarks for volumetric organ segmentation. However, their performance on tumors remains suboptimal, largely due to irregular tumor shapes, low contrast in medical scans, and diffuse boundaries. Even foundation models (FMs) trained on 3D medical images from scratch \cite{3DSAM1, 3DSAM2} show limited performance due to insufficient volumetric tumor annotations for pre-training and the high variability in tumor size, texture, and shape compared to organs. Addressing these challenges, this work advances tumor-specific 2D-to-3D adaptation to achieve more precise volumetric delineation.

\subsection{2D-to-3D Foundation Model Adaptation}
Training new 3D foundation models from scratch requires a vast amount of 3D data, which is often scarce and computationally more expensive to process. This prohibitive cost of training 3D foundation models from scratch has spurred interest in adapting 2D pre-trained models for volumetric tasks. Specifically, the abundance of high-quality 2D data \cite{2ddataset1, 2ddataset2, 2ddataset3} has driven rapid innovation in 2D foundation models, prompting efforts to adapt them for volumetric tasks. One prominent strategy involves injecting 3D adapters into a frozen 2D image encoder \cite{SAM-Adapter3, SAM-Adapter1}. By inflating 2D convolutions, these methods capture volumetric information with minimal trainable parameters. Analogous to image-to-video adaptation, \cite{SAM-Adapter2} uses space-depth adapters to incorporate 3D context. These approaches confirm that strong 3D performance is achievable by reusing the extensive knowledge embedded in 2D backbone weights. In medical imaging, such adaptation is particularly advantageous, where data heterogeneity, scanner variations, and high annotation costs complicate end-to-end 3D training.


\subsection{Visual-Language Alignment}
Studies have shown that visual-language models (VLM), FMs in other words, which embed semantic information into visual learning, surpass vision-only models in tasks like classification \cite{clipclassification}, segmentation \cite{clipsegmentation}, and anomaly detection \cite{clipad}. CLIP \cite{clip}, pre-trained on vast image-text datasets, achieves a good generalization ability in classification tasks, even effectively identifying previously unseen samples. Specifically in medical AI, MedCLIP \cite{medclip}, trained on extensive medical image-text data, excels in zero-shot prediction, supervised classification, and image-text retrieval. While \cite{clipseg2D1} and \cite{clipseg2D2} fine-tuned CLIP for 2D segmentation via cross-modal attention, recent work \cite{clipseg3D} extended this to 3D by fusing CLIP’s embeddings with 3D CNN features, but retained a modality-specific design where text guidance is applied only during decoding. In contrast, we propose \textbf{hierarchical vision-language prompting} that injects CLIP’s semantic priors directly into SAM’s encoder, enabling earlier fusion of anatomical context. This approach differs fundamentally from prior VLMs, which treat text as a post-hoc classifier, and aligns with clinical workflows where radiological reports guide attention during image interpretation.

\section{Methodology}
\label{sec:approach}

\subsection{Preliminaries}
\label{subsec:preliminaries}
We build upon two key components: SAM \cite{sam} and 3D SAM Adapter \cite{SAM-Adapter1}. SAM is a versatile, promptable segmentation model that that leverages a ViT-based \cite{ViT} encoder-decoder architecture and a prompt encoder, allowing interactive segmentation via points, boxes, or masks. Although pre-trained on extensive 2D image datasets, SAM faces two primary challenges in medical imaging: (1) the lack of domain-specific data in its pre-training leads to domain shift and reduced robustness, and (2) its 2D-centric design struggles to capture volumetric structure, typical of medical scans. 

To efficiently apply SAM to 3D tumor segmentation, Gong \textit{et al.} \cite{SAM-Adapter1} introduced the \textbf{3D SAM Adapter}, retaining SAM’s pre-trained encoder weights while converting its 2D operations into 3D. Specifically, 2D convolutions are replaced by 3D convolutions, and an additional dimension is added to patch embeddings, positional encodings, and attention queries, forming a 3D ViT. A lightweight spatial adapter, comprising projection and 3D convolution layers, is inserted between consecutive attention blocks to facilitate 3D feature learning. Only the convolution, spatial adapter, and normalization parameters are fine-tuned, while all other parameters remain frozen—ensuring memory efficiency and mitigating catastrophic forgetting. A multi-layer aggregation decoder \cite{multiagg} concatenates the encoder’s intermediate outputs to produce segmentation masks. Despite these adaptations, the 3D SAM Adapter \cite{SAM-Adapter1} supports only point-based prompts, which contain limited spatial context~\cite{sampler} to align prompt and image embeddings. Therefore, accuracy can still suffer in volumetric segmentation tasks.

To overcome these limitations, we propose \textbf{TAGS}, a 3D tumor-adaptive guidance framework that leverages multiple prompt types for improved tumor segmentation. \cref{fig2} illustrates the overall architecture for our proposed framework. 



\subsection{Annotation-free Organ Prompt} 
\label{subsec:organ-prompt}
Tumor segmentation is complicated by variability in size, location, and appearance. However, tumors generally reside within specific organs. Rather than multitask training for simultaneous organ and tumor segmentation \cite{multitaskseg1, multitaskseg2}, we propose a simpler method that circumvents the need for organ ground-truth labels.


As depicted in the top-left of \cref{fig2}, we use an automated tool, \textit{TotalSegmentator} \cite{totalsegmentator}, to generate pseudo-organ masks $M \in \mathbb{R}^{d \times h \times w}$, where $d$, $h$, and $w$ are the volume (input) dimensions. These masks are typically more accurate and less costly to obtain than tumor annotations. To align with SAM’s three-channel expectation, we replicate the original volume and replace its third channel with the organ mask, forming $I \in \mathbb{R}^{3 \times d \times h \times w}$. This overlay provides organ-level spatial context, guiding the model’s attention to relevant anatomical regions and improving tumor feature extraction.  

\subsection{Multi-level Feature Alignment}
\label{subsec:feature-alignment}
\textbf{Dual-category Text Prompt.} Since tumors often occupy $<$1\% of the entire organ volume, relying solely on a single “tumor” description is insufficient.  Inspired by  \cite{winclip} and \cite{medwinclip}, we adopt a hierarchical dual-category text prompting strategy that encodes both foreground (tumor) and background (healthy tissue). Each prompt includes high-level ``state" descriptors (e.g., ``lesion present/absent”) and generic ``template” descriptions, as shown in \cref{fig2}. Only the organ’s name is required, reducing dependence on specialized medical text encoders. To minimize bias, we generate multiple text prompts for each category (see supplementary material). We then aggregate their features with CLIP’s \cite{clip} text encoder to yield
$F_{text} \in \mathbb{R}^{c \times 2}$, where $c$ is the feature dimension.

\noindent\textbf{Stage-wise Alignment Adapter.} SAM's image encoder comprises 12 attention blocks (\cref{fig2}) grouped into four stages (three blocks per stage). We insert a lightweight adapter $A_s$ between stages to align image and text features more granularly. Each alignment adapter is designed with a single fully connected layer to minimize the number of parameters. Each input volume $I'$ passes through four stages with the assistance of a 3D adapter to generate visual features. Intermediate features at each stage are denoted as $F_{s} \in \mathbb{R}^{P_d \times P_h \times P_w \times c}, s\in \{1, 2, 3, 4\}$. Here, $P_d$, $P_h$, and $P_w$ represent the number of patches in the depth, height, and width dimensions after patchifying the original image. The alignment adapter $A_s$ could be formulated as:

\vspace{-0.2cm}
\begin{equation}
    A_s(F_s) = \sigma(F_sW_s) \text{, } s\in \{1, 2, 3, 4\},
\end{equation}

\noindent where $\sigma(\cdot)$ is the activation function and $W_s$ indicates the parameter of $A_s$. We also adopt residual connection to enhance robustness. With a scaling factor $\lambda = 0.2$, the final output of stage $s$, denoted as $F_{s}^{'}$, is defined as:

\vspace{-0.5cm}
\begin{equation}
    F_{s}^{'} = \lambda A_s(F_s) + (1-\lambda) F_s \text{, } s\in \{1, 2, 3, 4\}.
\end{equation}

\noindent\textbf{Multi-modal Alignment Loss.} We incorporate tumor annotations into a multi-modal alignment loss. First, we calculate the \textit{cosine similarity} ($sim(\cdot, \cdot)$) for each adapter output $A_s(F_s)$. After interpolation and softmax, we obtain dense predictions. Given the severe foreground-background imbalance in volumetric tumor segmentation, we employ a focal loss \cite{focal} combined with dice loss \cite{monai}:

\vspace{-0.5cm}
\begin{align}
    l_{a} = \sum_{s=1}^{4}\{\frac{1}{2}l_{focal}[\phi(sim(A_s(F_s), F_{text}), y)] + \nonumber \\
    \frac{1}{2}l_{dice}[\phi(sim(A_s(F_s), F_{text}), y]\},
\end{align}

\noindent where $\phi$ refers to interpolation and \textit{softmax}, and $y$ is the tumor ground truth. Although these predictions are dense, the final segmentation still requires the decoder with point-based prompts. Experimental results in \cref{subsec:ablation} verify this.
\begin{table*}[htbp]
\renewcommand\arraystretch{1.2}
\centering
\vspace{-0.9cm}
\resizebox{1\linewidth}{!}
{
\normalsize
\setlength{\abovecaptionskip}{-0.2cm}   
\setlength{\arrayrulewidth}{0.05em}
\begin{tabular}{ll|ccccccc}
\specialrule{0.10em}{0pt}{1pt}
\multirow{2}{*}{\textbf{Category}}                                            & \multirow{2}{*}{\textbf{Method}}   & \multicolumn{2}{c}{\textbf{Kidney Tumor}} & \multicolumn{2}{c}{\textbf{Liver Tumor}} & \multicolumn{2}{c}{\textbf{Pancreas Tumor}} & \multirow{2}{*}{\begin{tabular}[c]{@{}c@{}}\textbf{Tunable} \\ \textbf{Parameters}\end{tabular}}\\
&  & \textbf{Dice (\%)}   & \textbf{NSD (\%)}   & \textbf{Dice (\%)}  & \textbf{NSD (\%)}   & \textbf{Dice (\%)}    & \textbf{NSD (\%)} &    \\
\specialrule{0.05em}{1pt}{1pt}
\multirow{4}{*}{\begin{tabular}[c]{@{}c@{}}Classic\end{tabular}}            & nnUNet \cite{nnunet}           & 50.24           & 47.62          & 58.95          & 57.58          & 36.46            & 43.18           & 30.76 M\\
& 3D UX-Net \cite{3duxnet}     & 53.40           & 57.02          & 47.95          & 61.14          & 32.11            & 48.46           & 53.01 M\\
& Swin-UNETR \cite{swinunetr}  & 49.83           & 54.06          & 43.90          & 61.39          & 35.61            & 53.96           & 62.19 M\\
& UNETR++ \cite{unetr++}       & 54.15           & 55.09          & 35.33          & 47.42          & 21.50            & 37.22           & 55.70 M\\
\specialrule{0.05em}{0.5pt}{0.5pt}
\multirow{2}{*}{2D SAM-based}                    & SAM-B \cite{sam}             & 4.35           & 4.34          & 2.96          & 3.49          & 4.75            & 5.65           & -\\
& SAM-Med2D \cite{sammed2d}    & 24.57           & 29.86          & 12.43          & 17.70          & 17.78            & 27.15           & -\\
\specialrule{0.05em}{0.5pt}{0.5pt}
\multirow{4}{*}{3D SAM-based}                    & SAM-Med3D \cite{3DSAM1}      & 27.32           & 36.49          & 10.37          & 17.01          & 16.28            & 21.90           & -\\
& SAM-Med3D Turbo \cite{3DSAM1}& 47.83           & 66.40          & 19.34          & 30.95          & 34.64            & 57.62           & -\\
& SegVol \cite{3DSAM2}         & 27.44           & 30.33          & 40.92          & 31.72          & 54.10            & 64.56           & -\\
& SegVol w zoom \cite{3DSAM2}  & 28.14           & 31.51          & \textbf{66.93} & 64.62          & 58.60            & 66.03           & -\\
\specialrule{0.05em}{0.5pt}{0.5pt}
\multirow{1}{*}{CLIP-based}                    & Universal \cite{clipseg3D}      & 74.95           & 73.50          & 61.87          & 60.72          & \textbf{61.28}            & 70.51           & -\\
\specialrule{0.05em}{1pt}{1pt}
\multirow{4}{*}{\begin{tabular}[c]{@{}c@{}}2D-to-3D\end{tabular}}             & 3D Adapter (1pts) \cite{SAM-Adapter1}        & 76.67           & 81.95          & 52.80          & 69.46          & 55.02            & 77.92           & 25.46 M\\
& \textbf{TAGS (1pts)}      & \underline{80.39}    & \underline{87.69}   & 59.69          & 72.83          & 59.96     & \underline{82.05}    & 27.82 M\\
& 3D Adapter (3pts) \cite{SAM-Adapter1}        & 77.16           & 82.76          & 58.96          & \underline{73.46}         & 55.73            & 78.24           & 25.46 M\\
& \textbf{TAGS (3pts)}      & \textbf{80.83}  & \textbf{88.26} & \underline{66.23}    & \textbf{79.33} & \underline{61.04} & \textbf{83.10}   & 27.82 M\\
\specialrule{0.10em}{1pt}{0pt}
\end{tabular}
}
\normalsize
\caption{The comparison experiments between TAGS and other benchmarks. For SAM-based foundation models, we use them directly for inference, while other models are trained on the training set. The best results are \textbf{bold} and the second best ones are \underline{underlined}.}
\label{tab1}
\vspace{-0.2cm}
\end{table*}

\subsection{Generating Final Segmentations}
\label{subsec:obtain-result}
We follow the prompt encoder and decoder structure of \cite{SAM-Adapter1}. During training, we randomly sample $n=10$ points from foreground and background if a tumor is present; otherwise, all $n=10$ points come from the background. The decoder concatenates intermediate encoder outputs $F_{s}^{'}$ with the model input $I'$ to produce the segmentation $\hat{y}$. We optimize a dice loss on $\hat{y}$:
\begin{equation}
    L = l_{dice}(\hat{y}, y) + l_a .
\end{equation}
At inference, we preserve the organ prompt and alignment adapters but \textbf{omit} the text prompt. We randomly pick \underline{1 or 3} points inside the tumor region as input to the prompt encoder. The image is cropped around these points to match training patch sizes, and the encoder-decoder stack outputs the final mask. Experimental results in \cref{subsec:ablation} show that text-based alignment in training, combined with targeted prompts at test time, significantly enhances tumor segmentation accuracy. One to three point prompts balance usability and accuracy in volumetric segmentation. Using only one point offers a quick, minimal input requirement, making the approach more feasible in busy clinical settings. Meanwhile, adding two additional points (for a total of three) provides more geometric coverage within the tumor region, improving localization and boundary delineation. Empirical testing demonstrated that beyond three points, gains in segmentation accuracy diminish while user effort grows, so we chose to optimize performance without overburdening the prompt-creation process.

\section{Experiments and Results}
\label{sec:experiments}

\subsection{Setup}
\label{subsec:setup}
\textbf{Datasets and Evaluation Metrics.} We focus exclusively on tumor segmentation—arguably the most challenging task for SAM-based methods in medical imaging. Three public datasets, each targeting a different tumor type, are used: (a) the kidney tumor segmentation challenge dataset (KiTS21) \cite{kits21}; (b) the liver tumor segmentation benchmark dataset (LiTS) \cite{lits}; and (c) the pancreas tumor segmentation branch of the MSD challenge dataset (MSD-Pancreas) \cite{msd}. Each dataset is divided into training, validation, and test sets with a ratio of 70\%:10\%:20\%. We use Dice Score and Normalized Surface Dice (NSD) as evaluation metrics to assess both overall accuracy and tumor boundary precision by convention.


\noindent\textbf{Implementation Details.} We adopt \underline{SAM-B} as our image encoder (other SAM models are also used for benchmarking). To ensure consistent intermediate feature dimensions and align textual and visual features, we use CLIP ViT-L/14 as the text encoder. Data augmentation includes random rotations, flips, zooms, and intensity shifts. Foreground and background patches are randomly sampled at a 2:1 ratio during training. All experiments are conducted on a single NVIDIA A100 GPU using AdamW optimizer with the learning rate of $1e^{-4}$ and the batch size of 1 for 200 epochs.

\subsection{Comparison with State-of-the-Art Methods}
\label{subsec:comparison}
We benchmark our \textbf{TAGS} method against ten leading 3D medical segmentation techniques, including classical baselines (nnUNet \cite{nnunet}, Swin-UNETR \cite{swinunetr}, UNETR++ \cite{unetr++}, 3D UX-Net \cite{3duxnet}), 2D SAM-based approaches (SAM-B \cite{sam}, SAM-Med2D \cite{sammed2d}), 3D SAM-based and CLIP-based medical foundation models (SAM-Med3D \cite{3DSAM1}, SegVol \cite{3DSAM2}, and Universal Model \cite{clipseg3D}), and the original 3D SAM Adapter \cite{SAM-Adapter1}. For SAM-based models pre-trained on the three evaluation datasets, we perform direct inference on the test sets. By contrast, other models, including \textbf{TAGS}, are first trained on the respective training partitions. Table \ref{tab1} summarizes the comparative results.


\begin{figure*}[htbp]
\centering
\vspace{-0.8cm}
\setlength{\abovecaptionskip}{-0.1cm}   
\setlength{\belowcaptionskip}{-0.2cm}   
\centerline{\includegraphics[height=11cm]{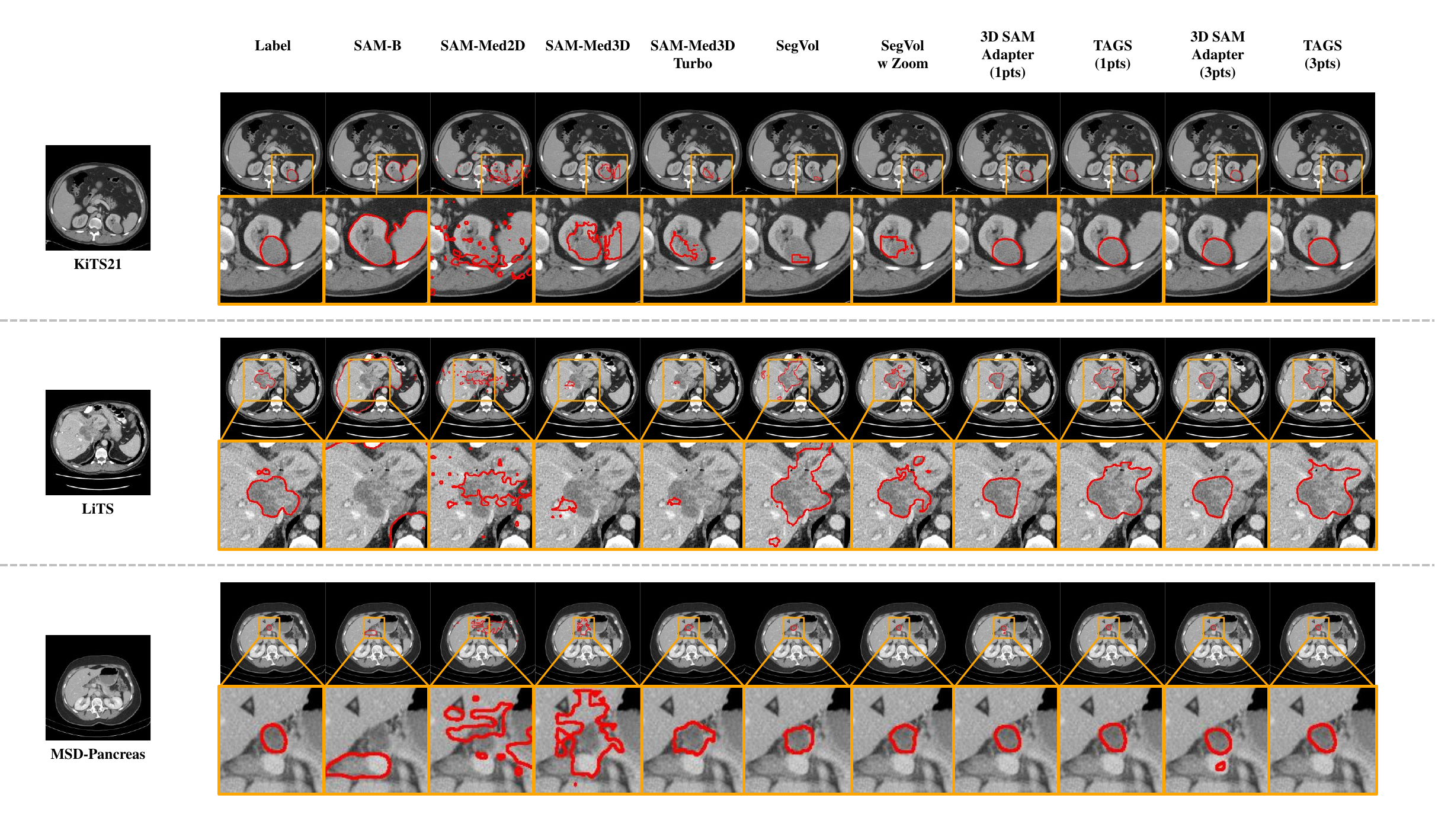}}
\caption{Qualitative visualizations of TAGS and SAM-based approaches for kidney, liver, and pancreas tumor segmentation. Lesion areas are highlighted with bounding boxes and zoomed in for detail.}
\label{fig3}
\end{figure*}

\noindent\textbf{Comparison with Classic Benchmarks.} TAGS exhibits substantial Dice score gains (exceeding 45\% improvement) over both CNN-based \cite{nnunet, 3duxnet} and transformer-based \cite{swinunetr, unetr++} methods for kidney and pancreas tumors. In liver tumor segmentation, TAGS still surpasses nnUNet \cite{nnunet} in Dice score, but the margin is smaller, possibly because multiple lesions in the liver require more than a single point prompt. Switching to three prompt points yields a considerable performance boost. TAGS also achieves a 25\% NSD advantage over nnUNet, highlighting its effectiveness in resolving complex tumor boundaries.

\noindent\textbf{Comparison with SAM- and CLIP-based Methods.} Table \ref{tab1} and \cref{fig3} show that \textbf{TAGS} surpasses both slice-by-slice and 3D SAM- and CLIP-based solutions, including SAM-Med3D \cite{3DSAM1}, SegVol \cite{3DSAM2}, and Universal Model \cite{clipseg3D}, despite their pre-training on these target datasets. Although SegVol’s “zoom-out-zoom-in” achieves a high Dice for LiTS and Universal Model performs well on MSD, \textbf{TAGS} reaches comparable performance with just two extra point prompts. Notably, TAGS maintains superior NSD, confirming its robust boundary delineation.  

\noindent\textbf{Comparison with 3D Adapters.} \textbf{TAGS} outperforms the 3D SAM Adapter \cite{SAM-Adapter1} by up to 13\% in Dice and 8\% in NSD, all while retaining a similar parameter count. This gain stems from TAGS’s more nuanced capture of small or irregular tumor features (\cref{fig3}). By embedding both \textbf{organ} and \textbf{text} prompts, \textbf{TAGS} ensures accurate segmentation with fewer interactive prompts, paving the way for broader clinical adoption.


\subsection{Ablation Studies}
\label{subsec:ablation}

We conduct comprehensive ablation studies to evaluate the design of our framework from various aspects. All experiments use a single-point prompt unless otherwise specified. Table \ref{tab2} reveals the effectiveness of \textit{organ prompt} and \textit{text prompt alignment}. Extended evaluations are detailed below.

\begin{table}[t]
\renewcommand\arraystretch{1.3}
\setlength{\belowcaptionskip}{-0.5cm}   
\resizebox{0.95\linewidth}{!}
{
\begin{tabular}{cc|cccccc}
\specialrule{0.10em}{0pt}{1pt}
\multirow{2}{*}{\begin{tabular}[c]{@{}c@{}}\textbf{Organ}\\ \textbf{Prompt}\end{tabular}} & \multirow{2}{*}{\begin{tabular}[c]{@{}c@{}}\textbf{Text Prompt}\\ \textbf{Alignment}\end{tabular}} & \multicolumn{2}{c}{\textbf{KiTS}} & \multicolumn{2}{c}{\textbf{LiTS}} & \multicolumn{2}{c}{\textbf{MSD-Pancreas}} \\
                                                                            &                                                                                  & \textbf{Dice $_\%$}           & \textbf{NSD $_\%$}           & \textbf{Dice $_\%$}          & \textbf{NSD $_\%$}           & \textbf{Dice $_\%$}            & \textbf{NSD $_\%$}            \\ 
\specialrule{0.05em}{0.5pt}{0.5pt}
    -                                                                       & -                                                                                & 76.67           & 81.95          & 52.80          & 69.46          & 55.02            & 77.92           \\
\checkmark                                                              & -                                                                                & 78.59           & 85.81          & 56.67          & 71.38          & 55.34            & 77.78           \\
    -                                                                       & \checkmark                                                                       & 78.56           & 85.04          & 57.22          & 71.42          & 58.34            & 80.32           \\
\checkmark                                                              & \checkmark                                                                       & \textbf{80.39}  & \textbf{87.69} & \textbf{59.69} & \textbf{72.83} & \textbf{59.96}   & \textbf{82.05} \\
\specialrule{0.10em}{1pt}{0pt}
\end{tabular}
}
\caption{Ablation studies of our framework. The best outcomes are highlighted in \textbf{bold}.}
\label{tab2}
\end{table}

\begin{table*}[t]
\vspace{-0.3cm}
\centering
\begin{minipage}[c]{0.48\textwidth}
\renewcommand\arraystretch{1.3}
    \resizebox{0.95\linewidth}{!}
    {
    \begin{tabular}{c|cccccc}
    \specialrule{0.10em}{0pt}{1pt}
    \multirow{2}{*}{\begin{tabular}[c]{@{}c@{}}\textbf{Organ Prompt}\\ \textbf{Type}\end{tabular}} & \multicolumn{2}{c}{\textbf{KiTS}} & \multicolumn{2}{c}{\textbf{LiTS}} & \multicolumn{2}{c}{\textbf{MSD-Pancreas}} \\
                                                                                 & \textbf{Dice $_\%$}            & \textbf{NSD $_\%$}            & \textbf{Dice $_\%$}           & \textbf{NSD $_\%$}            & \textbf{Dice $_\%$}             & \textbf{NSD $_\%$}             \\
    \specialrule{0.05em}{0.5pt}{0.5pt}
    \textbf{TotalSegmentator}                                                             & 80.39           & 87.69          & 59.69          & 72.83          & 59.96            & 82.05           \\
    \textbf{Ground Truth}                                                                 & 83.53           & 89.36          & 59.78          & 75.28          & 61.32            & 82.53           \\
    \specialrule{0.10em}{1pt}{0pt}
    \end{tabular}
    }
    \caption{Comparison results of using totalsegmentator-generated organ mask and ground truth mask.}
    \label{tab3}

    \vspace{0.3cm}
    \setlength{\belowcaptionskip}{-0.3cm}   
    \resizebox{0.95\linewidth}{!}
    {
    \renewcommand\arraystretch{1.3}
    \setlength{\abovecaptionskip}{-1cm}   
    \begin{tabular}{c|cccccc}
    \specialrule{0.10em}{0pt}{1pt}
    \multirow{2}{*}{} & \multicolumn{2}{c}{\textbf{KiTS}} & \multicolumn{2}{c}{\textbf{LiTS}} & \multicolumn{2}{c}{\textbf{MSD-Pancreas}} \\
                      & \textbf{Dice $_\%$}        & \textbf{NSD $_\%$}        & \textbf{Dice $_\%$}        & \textbf{NSD $_\%$}        & \textbf{Dice $_\%$}            & \textbf{NSD $_\%$}            \\
    \specialrule{0.05em}{0.5pt}{0.5pt}
    1st        & 70.51       & 78.50      & 49.69       & 65.79      & 47.75           & 71.92          \\
    2nd        & 76.13       & 84.25      & 54.32       & 70.14      & 54.04           & 76.97          \\
    3rd        & 78.90       & 85.98      & 58.37       & 72.08      & 55.89           & 77.72          \\
    4th        & 80.34       & 86.59      & 58.50       & 72.50      & 57.34           & 78.41          \\
    \specialrule{0.05em}{0.5pt}{0.5pt}
    \textbf{Whole Structure}               & \textbf{80.39}       & \textbf{87.69}      & \textbf{59.69}       & \textbf{72.83}      & \textbf{59.96}           & \textbf{82.05}          \\
    \specialrule{0.10em}{1pt}{0pt}
    \end{tabular}
    }
    \caption{Comparison of using aligned feature for segmentation.}
    \label{tab4}
\end{minipage}
\hspace{0.3cm}
\begin{minipage}[c]{0.48\textwidth}
\renewcommand\arraystretch{1.3}
    \resizebox{0.95\linewidth}{!}
    {
    \setlength{\abovecaptionskip}{-1cm}   
    \begin{tabular}{c|cccccc}
    \specialrule{0.10em}{0pt}{1pt}
    \multirow{2}{*}{\textbf{Text Encoder}} & \multicolumn{2}{c}{\textbf{KiTS}} & \multicolumn{2}{c}{\textbf{LiTS}} & \multicolumn{2}{c}{\textbf{MSD-Pancreas}} \\
                                                                            & \textbf{Dice $_\%$}        & \textbf{NSD $_\%$}        & \textbf{Dice $_\%$}        & \textbf{NSD $_\%$}        & \textbf{Dice $_\%$}            & \textbf{NSD $_\%$}            \\
    \specialrule{0.05em}{0.5pt}{0.5pt}
    \textbf{CLIP VLT-L/14}                                                           & 80.39       & 87.69      & 59.69       & 72.83      & 59.96           & 82.05          \\
    \textbf{MedCLIP}                                                                 & 79.15       & 86.23      & 57.66       & 72.87      & 56.70           & 76.96          \\
    \specialrule{0.10em}{1pt}{0pt}
    \end{tabular}
    }
    \caption{Ablation results between using original CLIP text encoder and medical-specific MedCLIP text encoder.}
    \label{tab5}

    \vspace{0.3cm}
    \setlength{\belowcaptionskip}{-0.3cm}   
    \resizebox{0.95\linewidth}{!}
    {
    \renewcommand\arraystretch{1.3}
    \setlength{\abovecaptionskip}{-1cm}   
    \begin{tabular}{l|cccccc}
    \specialrule{0.10em}{0pt}{1pt}
    \multicolumn{1}{c|}{\multirow{2}{*}{\begin{tabular}[c]{@{}c@{}}\textbf{Text Prompt}\\ \textbf{Type}\end{tabular}}} & \multicolumn{2}{c}{\textbf{KiTS}} & \multicolumn{2}{c}{\textbf{LiTS}} & \multicolumn{2}{c}{\textbf{MSD-Pancreas}} \\
    \multicolumn{1}{c|}{}                                                                            & \textbf{Dice $_\%$}        & \textbf{NSD $_\%$}        & \textbf{Dice $_\%$}        & \textbf{NSD $_\%$}        & \textbf{Dice $_\%$}            & \textbf{NSD $_\%$}            \\
    \specialrule{0.05em}{0.5pt}{0.5pt}
    \textbf{One-category}                                                                                        & 76.47       & 83.72      & 54.59       & 68.91      & 52.91           & 74.89          \\
    \specialrule{0.05em}{0.5pt}{0.5pt}
    \textbf{Dual-category}                                                                                        & 77.04       & 84.08      & 55.82      & 70.49      & 54.60           & 74.94          \\
    \textbf{+state}                                                                                           & 78.00       & 84.65      & 57.62       & 71.96      & 57.48           & 79.14          \\
    \textbf{+template}                                                                                        & 79.12       & 86.45      & 56.66       & 71.39      & 56.21           & 77.22          \\
    \textbf{+both}                                                                                            & \textbf{80.39}       & \textbf{87.69}      & \textbf{59.69}       & \textbf{72.83}      & \textbf{59.96}           & \textbf{82.05}          \\
    \specialrule{0.10em}{1pt}{0pt}
    \end{tabular}
    }
    \caption{Comparison results of using different text prompts.}
    \label{tab6}
\end{minipage}
\end{table*}

\noindent\textbf{Precision of Organ Prompt.} To avoid extra annotation costs, we utilize \textit{TotalSegmentator} \cite{totalsegmentator} to generate organ prompts. Since organ segmentation is relatively easier, \textit{TotalSegmentator} could produce high-quality organ masks even without fine-tuning. \textit{TotalSegmentator} achieves organ segmentation Dice scores of 88.96\% on KiTS, 88.57\% on LiTS, and 78.54\% on MSD-Pancreas, respectively. To examine whether more accurate organ prompts can further improve tumor segmentation performance, we also compare results using ground truth organ prompts for training and testing, as shown in Table \ref{tab3}. The results indicate that tumor segmentation performance could be boosted with more precise organ information.

\noindent\textbf{Effects of Dual-category Text Prompt.} Our text prompt design hinges on two key elements: the dual-category foreground and background distinction and the two-level description. Table \ref{tab6} proves that both are crucial for achieving optimal semantic alignment. For one-category text prompt, we use ``\{obj\} with tumor", where \{obj\} refers to the organ. For dual-category prompts without two-level descriptions, we use ``healthy \{obj\}" and ``\{obj\} with tumor" to describe each category. We confirm that including descriptions for both tumor and background improves semantic understanding. Diverse descriptions at the state and template levels also contributes significantly to performance gains.

To further assess the necessity of a medically fine-tuned text encoder, we compare the results of using MedCLIP \cite{medclip} and our original CLIP text encoder in Table \ref{tab5}. Both approaches perform similarly, though our original setup shows a slight advantage. We attribute this to the text prompt design, which avoids extensive medical priors (e.g., tumor texture, contours) and emphasizes clarity and generality in state-level and template-level descriptions. This design makes the prompts closer to natural language, enabling effective multimodal alignment without fine-tuning the text encoder on medical data. It offers a new way to integrate multimodal information for medical image analysis.

\begin{figure}[ht]
\centering
\setlength{\belowcaptionskip}{-0.1cm}   
\centerline{\includegraphics[height=3.4cm]{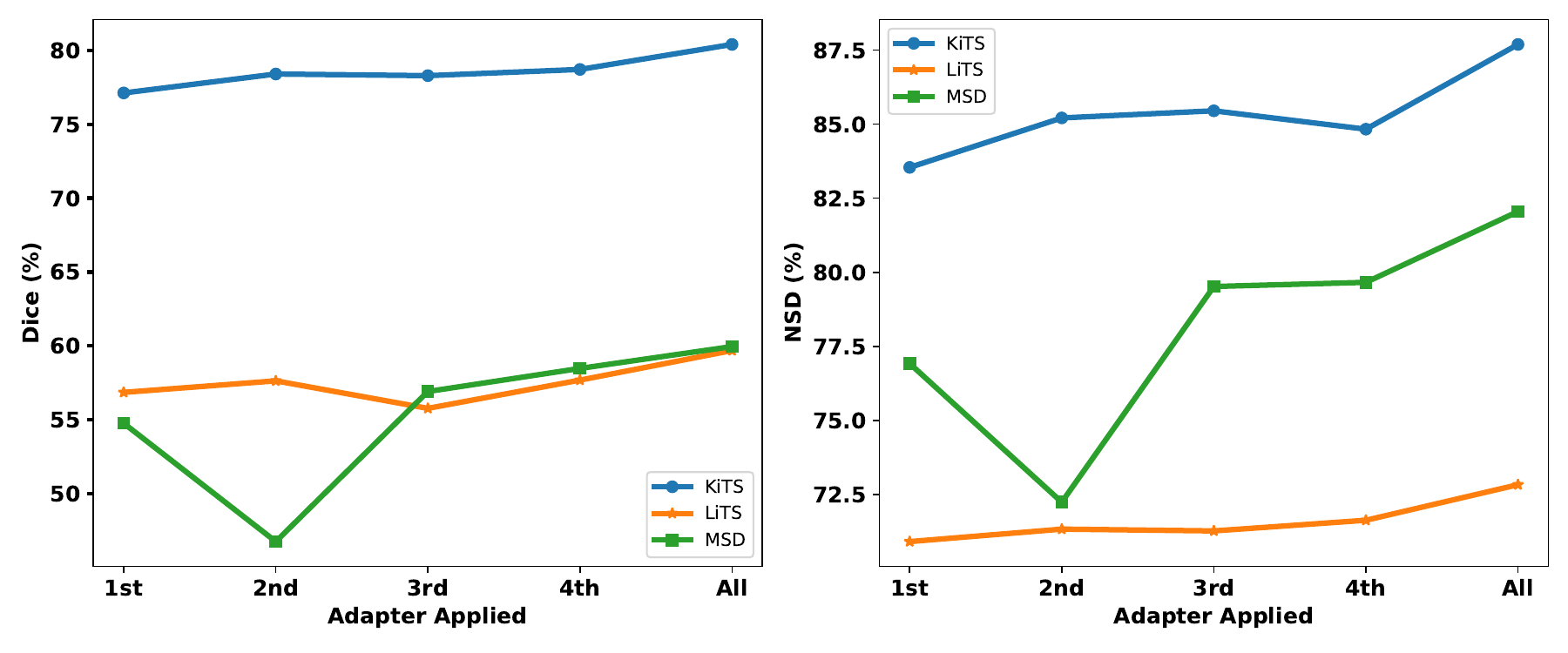}}
\caption{Effectiveness analysis of the necessity of using multi-level alignment adapters. When multi-level adaptation is not used, only one of the four adapters is employed.}
\label{fig4}
\end{figure}

\begin{figure}[ht]
\centering
\vspace{-0.25cm}
\setlength{\abovecaptionskip}{-0.05cm}   
\setlength{\belowcaptionskip}{-0.2cm}   
\centerline{\includegraphics[height=6.7 cm]{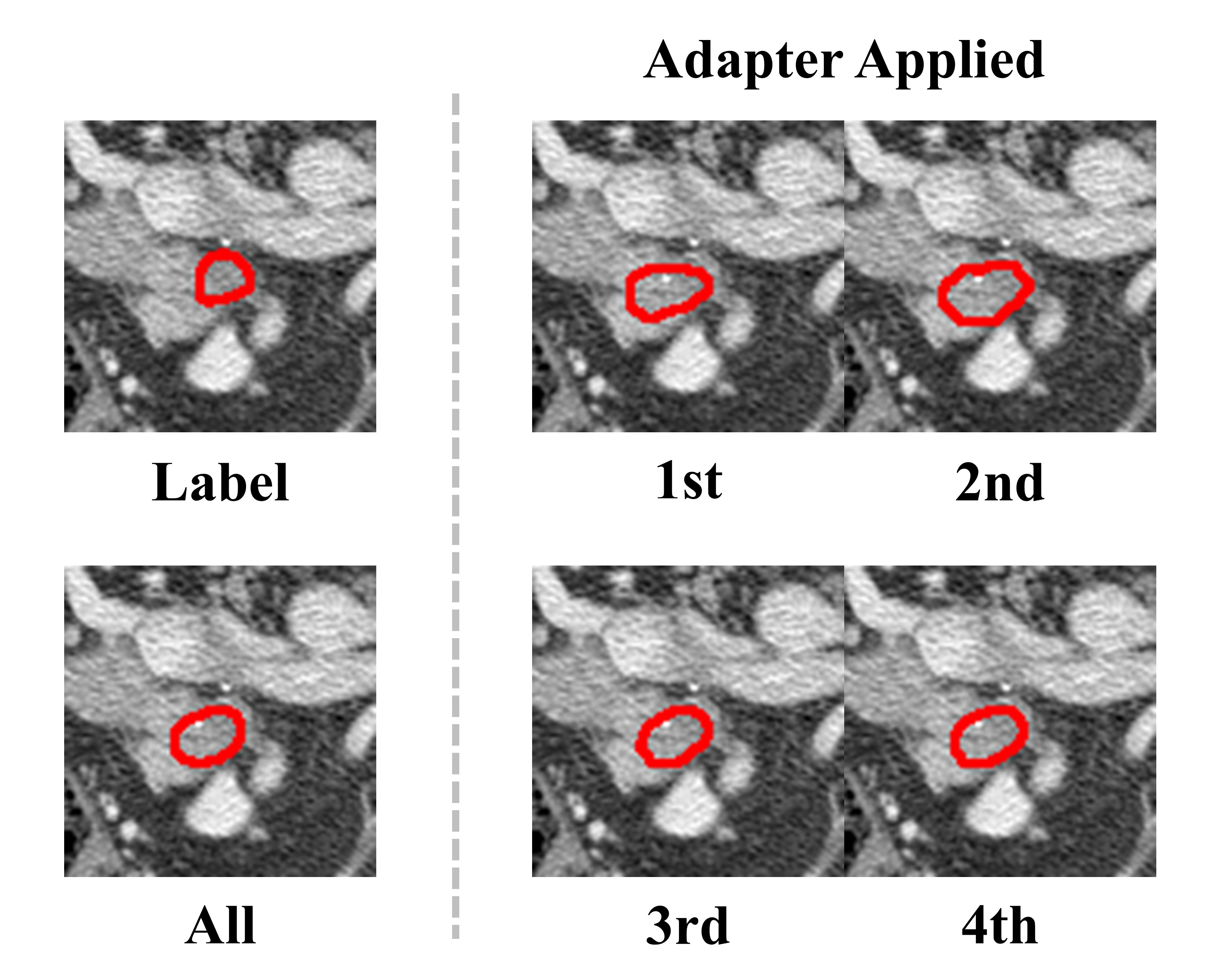}}
\caption{Visualization of segmentation results on the MSD-Pancreas dataset using different alignment strategies.}
\label{fig5}
\end{figure}

\noindent\textbf{Effects of Multi-level Alignment.} We first evaluate the necessity of using multi-level alignment adapters. In \cref{fig4}, we display the segmentation performance when training with each alignment adapter $A_s$ individually. Visualization results for each model on the MSD-Pancreas dataset are shown in \cref{fig5}. Among the single adapter approaches, models with alignment adapters placed in the third and fourth stages produce results closest to the segmentation ground truth. This aligns with standard foundational model adaptation practices where adapters are positioned near the end of encoders. However, the model using multi-level adapters achieves the highest Dice score and NSD values across all datasets. These results suggest that integrating semantic prompts at multiple stages of the image encoder provides a more refined approach than relying on a single adapter for alignment, allowing the model to learn from textual information more effectively.

We further test the image encoder’s ability to learn the spatial features of the image. In this test, we invoke a CLIP-like structure. We remove the prompt encoder and mask decoder while reintroducing the text encoder for feature alignment. Each alignment adapter's output feature $A_s(F_s)$ is aligned with the text feature $F_{text}$. Linear interpolation is then applied to match input dimensions, producing the segmentation output. We calculate the aligned prediction results for each $A_s$ separately and summarize these results alongside those from the full model structure in Table \ref{tab4}. We observed that as the model progresses through layers, the aligned prediction accuracy produced by adapters at different stages gradually improves. After aligned with the text encoder, predictions from the fourth alignment adapter show an average Dice score difference of 2.41\% and an NSD difference of 2.12\% compared to the full encoder-decoder model across three datasets. This reveals that the organ and text prompts have enabled the image encoder to learn the image information effectively. Yet, high-quality point prompts can further enhance segmentation accuracy.

\begin{table}[ht]
\renewcommand\arraystretch{1.3}
\setlength{\belowcaptionskip}{-0.1cm}   
\resizebox{0.95\linewidth}{!}
{
\begin{tabular}{ccccccc}
\specialrule{0.10em}{0pt}{1pt}
\multicolumn{1}{c|}{\multirow{2}{*}{\begin{tabular}[c]{@{}c@{}}\textbf{Selection}\\ \textbf{Strategy}\end{tabular}}} & \multicolumn{2}{c}{\textbf{KiTS}} & \multicolumn{2}{c}{\textbf{LiTS}} & \multicolumn{2}{c}{\textbf{MSD-Pancreas}} \\
\multicolumn{1}{c|}{}                                                                              & \textbf{Dice $_\%$}        & \textbf{NSD $_\%$}        & \textbf{Dice $_\%$}        & \textbf{NSD $_\%$}        & \textbf{Dice $_\%$}            & \textbf{NSD $_\%$}            \\
\specialrule{0.05em}{0.5pt}{0.5pt}
\multicolumn{1}{c|}{\textbf{random (1pts)}}                                                                 & 80.39       & 87.69      & 59.69       & 72.83      & 59.96           & 82.05           \\
\multicolumn{1}{c|}{\textbf{edge (1pts)}}                                                                   & 78.91       & 87.39      & 59.11       & 72.09      & 58.02           & 79.87          \\
\multicolumn{1}{c|}{\textbf{edge (3pts)}}                                                                   & 82.62       & 90.27      & 66.47       & 78.62      & 58.92           & 81.12          \\
\multicolumn{1}{c|}{\textbf{central (1pts)}}                                                                & 82.68       & 90.29      & 67.09       & 81.77      & 59.05           & 81.22          \\
\specialrule{0.05em}{0pt}{01pt}
\specialrule{0.05em}{01pt}{0pt}
\textbf{ICC (\%)}                                                                                           & 90.54       & 91.74      & 90.40       & 93.81      & 96.78           & 96.93   \\
\specialrule{0.10em}{1pt}{0pt}
\end{tabular}
}
\caption{Tumor segmentation performance with different point prompt selection strategies. 
 Intra-class Correlation Coefficient (ICC) of each strategy is reported in the last row.}
\label{tab7}
\end{table}

\begin{figure}[ht]
\centering
\setlength{\belowcaptionskip}{0 cm}   
\centerline{\includegraphics[height=6.2 cm]{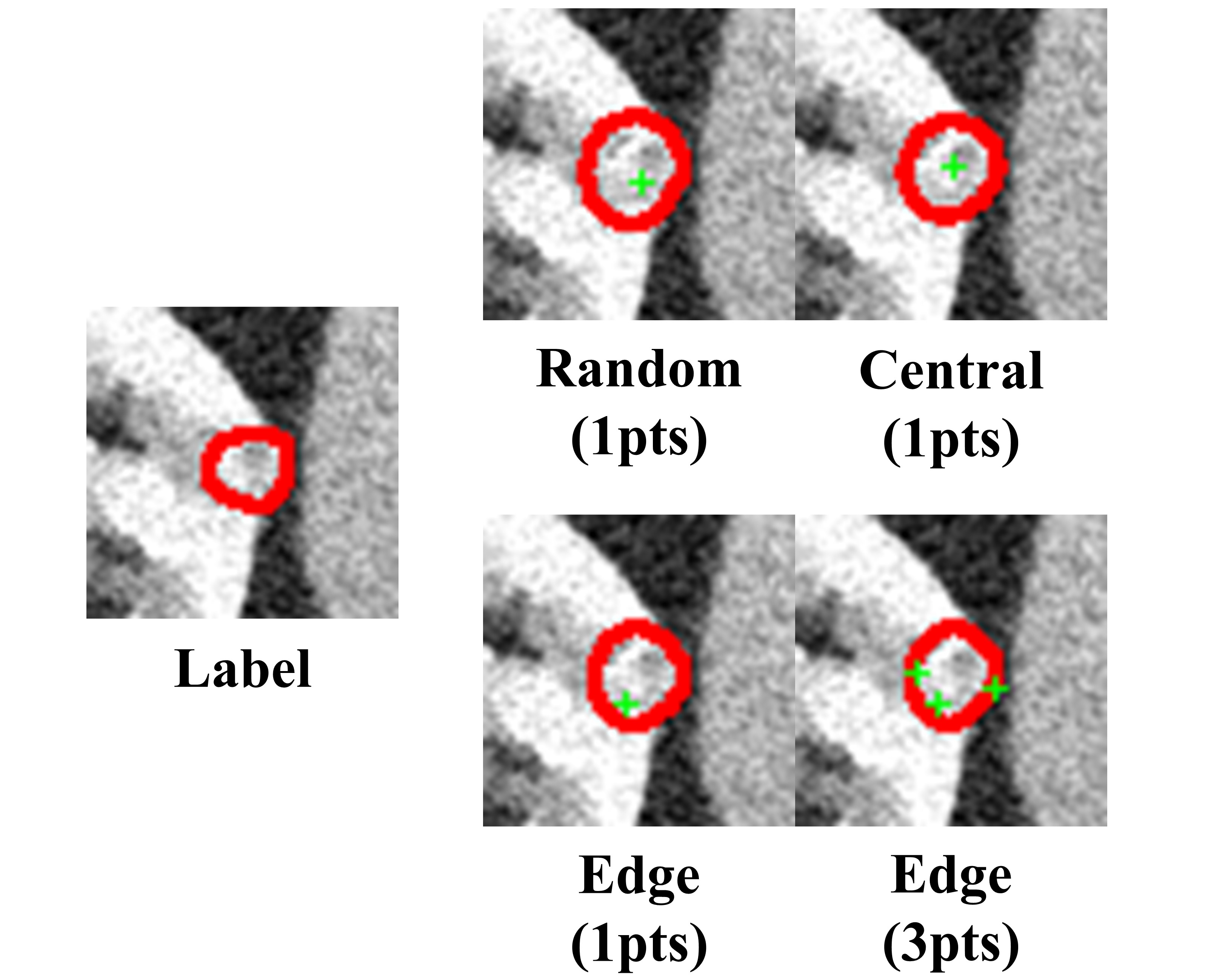}}
\caption{Illustration of different point prompt selection strategies. Green marks on each prediction indicate the selected points.}
\label{fig6}
\end{figure}

\noindent\textbf{Robustness Against Point Prompts.} To explore the robustness of TAGS under different point prompts, we examine how point prompts from various locations impact segmentation performance. In our original training setup, TAGS randomly selects coordinate points from the tumor region as point prompts during inference. Here, we compare the effects of two new point selection strategies on model performance: central selection and edge selection, as illustrated in \cref{fig6}. central selection uses the central point of the tumor area as the prompt, while edge selection randomly chooses points along the tumor’s boundary. For samples with multiple tumors, these operations are applied to the largest tumor only. Table \ref{tab7} presents the segmentation performance for each strategy, with edge selection tested using one-point and three-point scenarios. We find that both the quality and quantity of point prompts will impact model predictions. When prompt quality is low, such as providing only a single point on the edge, predictions for the tumor region can be more susceptible to interference. Increasing the number of edge prompts or using high-quality central point prompts leads to more accurate predictions. 

Interestingly, different point prompts yield only about a 3\% variation in segmentation outcomes, thanks to the precision of the proposed adapter. This indicates that, unlike other SAM-based methods, TAGS does not rely heavily on interactive prompts to achieve high-quality predictions. Might be an exception, on the LiTS dataset, using the tumor center point prompt and multiple edge prompts lead to approximately a 10\% improvement in both Dice score and NSD. We attribute this to the fact that these combined strategies allow the cropped inference patch around the prompt center to cover more of the tumor region. To further validate the robustness of TAGS with different point prompts, we also calculate the Intra-class Correlation Coefficient (ICC) for the predictions of the above strategies, as shown in Table \ref{tab7}. The results confirm a high consistency in model predictions across different point prompts, with ICC values for Dice score and NSD exceeding 90\% in each dataset.

\section{Conclusion}
\label{sec:con&dis}
Volumetric tumor segmentation remains a critical challenge in medical imaging due to the limitations of 2D foundation models like SAM in handling 3D spatial relationships and domain shifts. We introduce \textbf{TAGS}, a framework that bridges SAM’s 2D priors to 3D medical tasks through multi-prompt adaptation, achieving three key advancements: (1) annotation-free organ prompting that leverages automated segmentation tools to guide tumor localization without costly annotations, (2) hierarchical vision-language alignment that integrates CLIP’s semantic priors with SAM’s spatial acuity, and (3) parameter-efficient 3D adaptation via lightweight multi-stage adapters. 
Extensive evaluations on three public datasets (kidney, liver, and pancreatic tumors) confirm that \textbf{TAGS} substantially outperforms both SAM-based baselines and state-of-the-art 3D segmentation methods, including 3D foundation models specifically trained on medical data. Notably, TAGS delivers these gains with minimal additional parameters, offering a lightweight yet powerful solution for 2D-to-3D model adaptation. Future work could explore extending this multi-prompt strategy to other clinical tasks and integrating additional semantic cues, further advancing the domain-specific capabilities of foundation models in medical imaging. Code and models are released to foster reproducibility.

\section{Acknowledgement}
\label{sec:ack}
This research was supported by NIH under grant numbers R01-HL171376 and U01-CA268808.
{
    \small
    \bibliographystyle{ieeenat_fullname}
    \bibliography{main}
}
\clearpage
\setcounter{page}{1}
\maketitlesupplementary


\setcounter{table}{7}
\setcounter{figure}{6}
\setcounter{section}{0}

\section{Dataset Details}
\label{sec:dataset}

Table \ref{suptab1} presents detailed information on each dataset. The KiTS dataset \cite{kits21} originates from the MICCAI 2021 Kidney and Kidney Tumor Segmentation Challenge. The LiTS dataset \cite{lits} comes from the MICCAI 2017 Liver Tumor Segmentation Challenge, while the MSD-Pancreas dataset \cite{msd} is part of the 2018 Medical Segmentation Decathlon’s pancreas branch. Although all datasets initially included training and test splits, we re-divided the training sets into training, validation, and test subsets due to the lack of organ and tumor annotations in the original test sets. In both training and inference, we used only tumor labels and excluded organ annotations from the process.

\begin{table*}[htbp]
\renewcommand\arraystretch{1.5}
\setlength{\belowcaptionskip}{-0.1cm}   
\resizebox{0.95\linewidth}{!}
{
\normalsize
\begin{tabular}{c|ccccccc}
\specialrule{0.10em}{0pt}{1pt}
\textbf{Dataset} & \textbf{Train} & \textbf{Validation} & \textbf{Test} & \textbf{Resampled Spacing} & \textbf{Intensity Clipping Range} & \textbf{Patch Size}                                     & \textbf{Annotations Included}     \\
\specialrule{0.05em}{0.5pt}{0.5pt}
\specialrule{0.05em}{0.5pt}{0.5pt}
\textbf{KiTS}             & 209                       & 30                          & 61                    & (1, 1, 1)                  & {[}-52, 247{]}                    & 128 $\times$ 128 $\times$ 128 & Kidney, Kidney Tumor, Kidney Cyst \\
\textbf{LiTS}             & 83                        & 11                          & 24                    & (1, 1, 1)                  & {[}-17, 201{]}                    & 128 $\times$ 128 $\times$ 128 & Liver, Liver Tumor                \\
\textbf{MSD-Pancreas}     & 196                       & 28                          & 57                    & (1, 1, 1)                  & {[}-39, 204{]}                    & 128 $\times$ 128 $\times$ 128 & Pancreas, Pancreas Tumor          \\
\specialrule{0.10em}{1pt}{0pt}
\end{tabular}
}
\caption{Details of the three evaluation datasets.}
\label{suptab1}
\end{table*}

\begin{table*}[htbp]
\renewcommand\arraystretch{1.2}
\centering
\resizebox{0.9\linewidth}{!}
{
\normalsize
\setlength{\abovecaptionskip}{-0.1cm}   
\setlength{\arrayrulewidth}{0.05em}
\begin{tabular}{ll|cccccc}
\specialrule{0.10em}{0pt}{1pt}
\multirow{2}{*}{\textbf{Category}}                                            & \multirow{2}{*}{\textbf{Method}}   & \multicolumn{2}{c}{\textbf{Kidney Tumor}} & \multicolumn{2}{c}{\textbf{Liver Tumor}} & \multicolumn{2}{c}{\textbf{Pancreas Tumor}}\\
&  & \textbf{Dice (\%)}   & \textbf{NSD (\%)}   & \textbf{Dice (\%)}  & \textbf{NSD (\%)}   & \textbf{Dice (\%)}    & \textbf{NSD (\%)} \\
\specialrule{0.05em}{1pt}{1pt}
\multirow{1}{*}{2D SAM-based}
& SAM-Med2D \cite{sammed2d}    & 0.67           & 15.75          & 0.11          & 2.98          & 1.47            & 14.61           \\
\specialrule{0.05em}{0.5pt}{0.5pt}
\multirow{4}{*}{3D SAM-based}                    & SAM-Med3D \cite{3DSAM1}      & 67.84           & 81.33          & 36.33          & 51.34          & 60.39            & 80.64           \\
& SAM-Med3D Turbo \cite{3DSAM1}& 69.16           & 85.60          & 38.76          & 52.45          & \underline{60.65}            & 80.56           \\
& SegVol \cite{3DSAM2}         & 36.30           & 37.77          & 23.82          & 19.28          & 10.78            & 17.30           \\
& SegVol w zoom \cite{3DSAM2}  & 52.20           & 50.82          & 54.27 & 49.03          & 26.23            & 34.30           \\
\specialrule{0.05em}{1pt}{1pt}
\multirow{2}{*}{\begin{tabular}[c]{@{}c@{}}2D-to-3D\end{tabular}}
& \textbf{TAGS (1pts)}      & \underline{80.39}    & \underline{87.69}   & \underline{59.69}          & \underline{72.83}          & 59.96     & \underline{82.05}    \\
& \textbf{TAGS (3pts)}      & \textbf{80.83}  & \textbf{88.26} & \textbf{66.23}    & \textbf{79.33} & \textbf{61.04} & \textbf{83.10}   \\
\specialrule{0.10em}{1pt}{0pt}
\end{tabular}
}
\normalsize
\caption{The comparison experiments between TAGS and fine-tuned SAM-based benchmarks. The best results are \textbf{bold} and the second best ones are \underline{underlined}.}
\label{suptab2}
\end{table*}

\section{Supplementary Implementation Details}
\label{sec:implement}

\noindent\textbf{Data Processing.} Our pre-processing pipeline follows the approach in \cite{SAM-Adapter1}. We resample anisotropic images to the target spacing, followed by intensity clipping and normalization. For data augmentation, each sample has a 50\% probability of undergoing random flipping, rotation, or intensity shifting, and a 30\% probability of random zooming. Other settings have been detailed in Sec. {\color{iccvblue}{4.1}}.

\noindent\textbf{Text Prompt Design.} As described in Sec. {\color{iccvblue}{3.3}}, our text prompt incorporates multiple state-level and template-level descriptions for each category. \cref{supfig2} exhibits all prompts we employ at both levels, where \{obj\} refers to specific organ name, and \{c\} represents a state-level prompt. At state-level, we describe the tumor or background region using varied language. While at template-level, we employ clear, general sentences that include descriptions reflecting the data augmentation conditions for each sample. For each category, the embeddings generated from all descriptions by the CLIP text encoder are averaged to form a final embedding, guiding the image encoder’s understanding of image features. This approach effectively avoids potential biases from using a single text description.

\noindent\textbf{$\lambda$ Design.} $\lambda$ is set to 0.2, as it achieves optimal performance in our experiments. We fix it to avoid extra learnable parameters and ensure model efficiency.

\section{Additional Qualitative Evaluations}
\label{sec:quatitative}

\noindent\textbf{Qualitative Visualization on Other Benchmarks.} As a supplement to Fig. {\color{iccvblue}{3}}, we present a qualitative visualization comparison of TAGS with four classic volumetric segmentation benchmarks and CLIP-based benchmarks in \cref{supfig3}. Both Fig. {\color{iccvblue}{3}} and \cref{supfig3} visualize the middle slice within the non-zero pixel range of each volume along the depth dimension.

It can be seen that, although these benchmarks have succeeded in organ segmentation, their performance in tumor segmentation remains limited. All five models struggle to accurately capture tumor boundaries and shape information. For small pancreas tumors with high variability in shape and texture, nnUNet \cite{nnunet} and 3D UX-Net \cite{3duxnet} even show segmentation failures.

\begin{table}[ht]
\renewcommand\arraystretch{1.3}
\setlength{\belowcaptionskip}{-0.2cm}   
\resizebox{0.95\linewidth}{!}
{
\begin{tabular}{c|cccccc}
\specialrule{0.10em}{0pt}{1pt}
\multirow{2}{*}{\begin{tabular}[c]{@{}c@{}}\textbf{Text}\\ \textbf{Encoder}\end{tabular}} & \multicolumn{2}{c}{\textbf{KiTS}} & \multicolumn{2}{c}{\textbf{LiTS}} & \multicolumn{2}{c}{\textbf{MSD-Pancreas}} \\
& \textbf{Dice $_\%$}            & \textbf{NSD $_\%$}            & \textbf{Dice $_\%$}           & \textbf{NSD $_\%$}            & \textbf{Dice $_\%$}             & \textbf{NSD $_\%$}             \\
\specialrule{0.05em}{0.5pt}{0.5pt}
BERT \cite{bert}                                                            & 78.45           & 85.97          & 57.22          & 70.71          & 46.11            & 68.90           \\
MedCLIP \cite{medclip}                                                                 & 79.15           & 86.23          & 57.66          & \textbf{72.87}          & 56.70            & 76.96           \\
CT-CLIP \cite{ctclip}                                                                 & 80.24           & 86.82          & 59.21          & 73.23          & 55.72            & 77.86           \\
\specialrule{0.05em}{0.5pt}{0.5pt}
\textbf{TAGS}                                                                 & \textbf{80.39}           & \textbf{87.69}          & \textbf{59.69}          & 72.83          & \textbf{59.96}            & \textbf{82.05}           \\
\specialrule{0.10em}{1pt}{0pt}
\end{tabular}
}
\normalsize
\caption{Comparison of different text encoders. All experiments are conducted using a single-point prompt, with the best results highlighted in \textbf{bold}.}
\label{suptab3}
\end{table}

\begin{figure}[t]
\centering
\setlength{\belowcaptionskip}{-0.2cm}   
\centerline{\includegraphics[height=6cm]{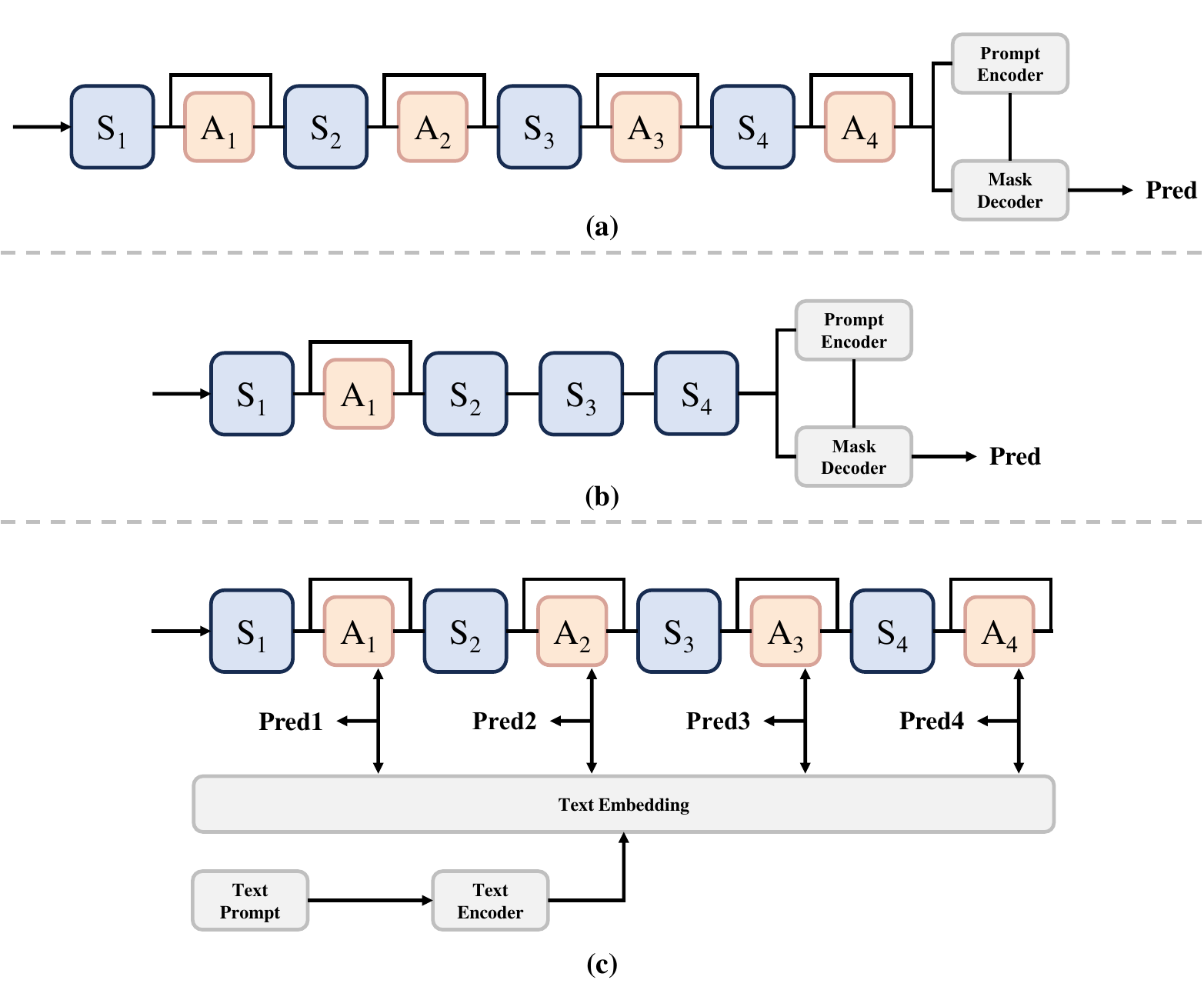}}
\caption{Illustration of different model structures: (a) TAGS when conducting inference; (b) single alignment adapter ablation; (c) directly utilizing aligned feature for prediction.}
\label{supfig1}
\end{figure}

\noindent\textbf{Comparison with Fine-tuned SAM-based Models.} For pretrained SAM-based medical frameworks, though we think the direct inference is a fair comparison (Sec. {\color{iccvblue}{4.2}}), we fine-tuned them on our training sets for further evaluation. As shown in Table \ref{suptab2}, each one consistently underperformed relative to TAGS. We also observe a performance decline in some methods after fine-tuning, which is likely due to the limited training set size being insufficient for effective fine-tuning.

\section{Additional Ablation Studies}
\label{sec:ablation}

\noindent\textbf{Model Structure of Multi-level Alignment Ablation.} To clarify the distinctions in Sec. {\color{iccvblue}{4.3}}, we present visual representations of the model structures in \cref{supfig1} for the ``single alignment adapter ablation experiment" referenced in Fig. {\color{iccvblue}{4}}, the ``CLIP-like structure that directly predicts using aligned features" detailed in Table {\color{iccvblue}{4}}, and the standard TAGS inference model structure.

\noindent\textbf{Utilizing Different Text Encoders.} Table \ref{suptab3} presents an evaluation of the effectiveness of different text encoders within the TAGS structure. We compare the original TAGS configuration, which uses CLIP, against alternatives such as BERT \cite{bert}, MedCLIP \cite{medclip}, and CT-CLIP \cite{ctclip} as text encoders. The results indicate that substituting CLIP with BERT-like text-only encoders leads to a decline in performance, falling below the no-text-embedding baseline shown in Table {\color{iccvblue}{2}}. Furthermore, replacing the text encoder with MedCLIP or CT-CLIP produces comparable results, with a performance difference of approximately 3\%, reinforcing our assertion that medical fine-tuning was unnecessary for this study.


\noindent\textbf{Additional Results for Robustness Against Point Prompts.} To further support the content in Table {\color{iccvblue}{7}}, we examine the performance differences between ``random selection" and ``edge selection" as point prompts increase, as detailed in \cref{supfig4}. For ``central selection", since the prompt point is always at the tumor center, increasing the number of prompts does not introduce new points. Therefore, it is not included in the discussion. We evaluated scenarios with 1, 3, 5, 7, and 10 point prompts. Results show that for ``random selection", increasing the number of points has little impact on performance. Meanwhile, for ``edge selection", segmentation accuracy stabilizes after 3 prompts, with minimal improvement from more points. This suggests that TAGS achieves strong performance with few prompts, underscoring its efficiency in learning image information and robustness against different point prompts.

\begin{figure*}[htbp]
\centering
\vspace{-0.3cm}
\setlength{\abovecaptionskip}{-0.1cm}   
\setlength{\belowcaptionskip}{-0.2cm}   
\centerline{\includegraphics[height=4.5cm]{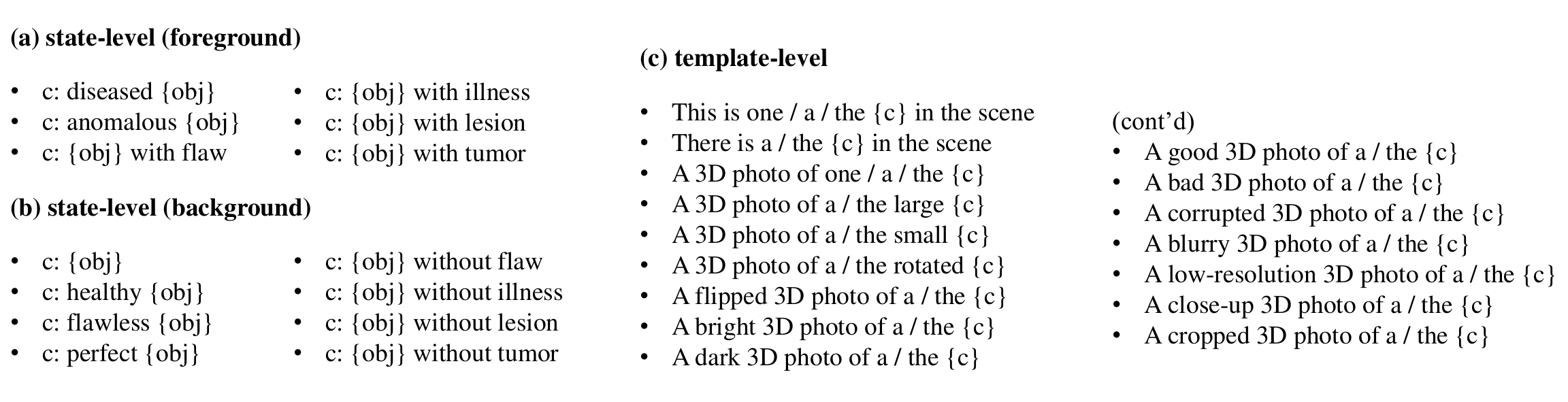}}
\caption{Lists of two-level text prompt description that we used for feature alignment. \{obj\} denotes the name of specific organ and \{c\} refers to a state-level description.}
\label{supfig2}
\end{figure*}

\begin{figure*}[htbp]
\centering
\vspace{0.5cm}
\setlength{\abovecaptionskip}{-0.1cm}   
\setlength{\belowcaptionskip}{-0.1cm}   
\centerline{\includegraphics[height=10cm]{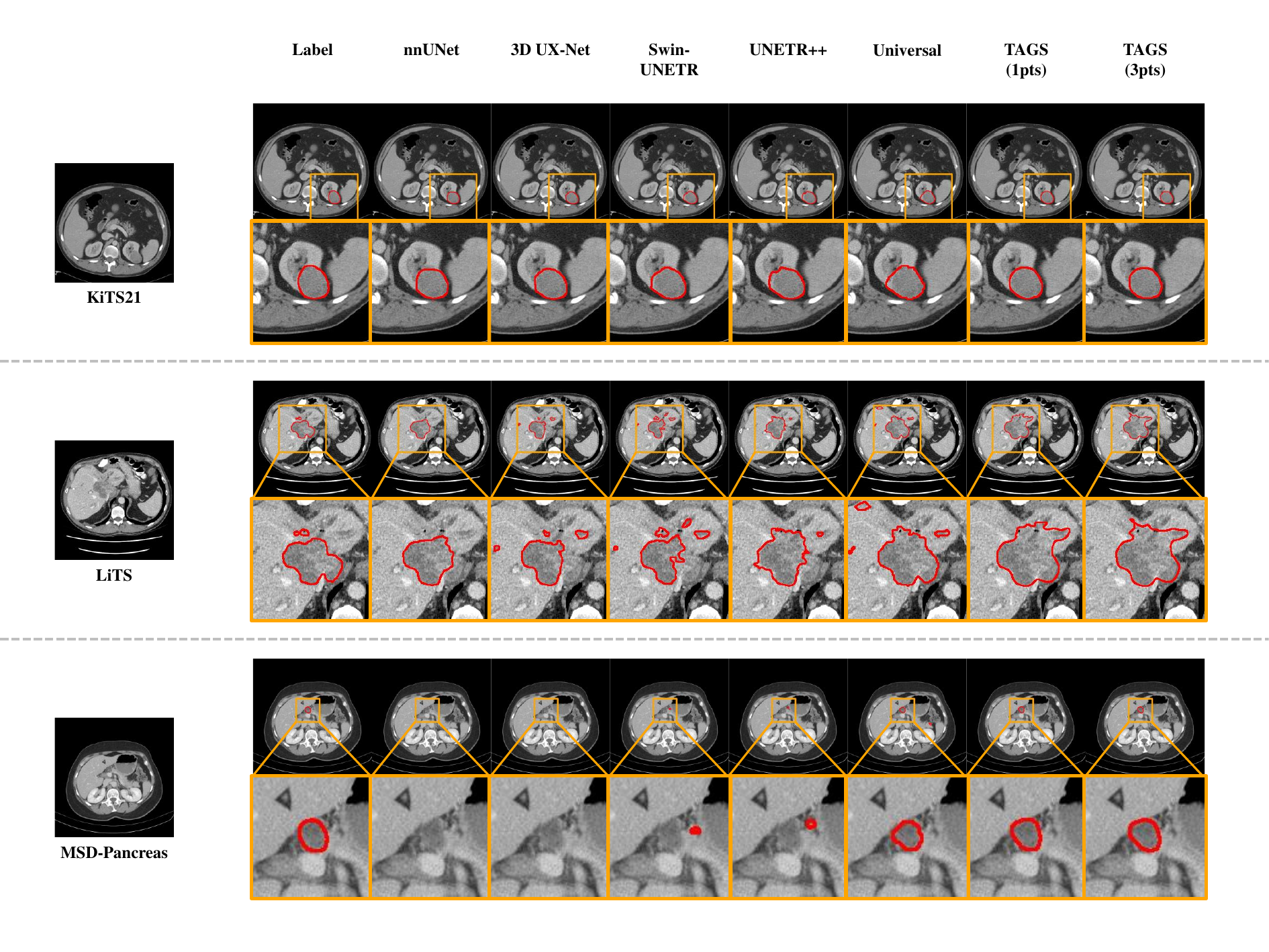}}
\caption{Qualitative visualizations of TAGS and other volumetric segmentation benchmarks approaches for kidney, liver, and pancreas tumor segmentation. Lesion areas are highlighted with bounding boxes and zoomed in for detail.}
\label{supfig3}
\end{figure*}

\begin{figure*}[h!]
\centering
\vspace{0.6cm}
\centerline{\includegraphics[height=3.5cm]{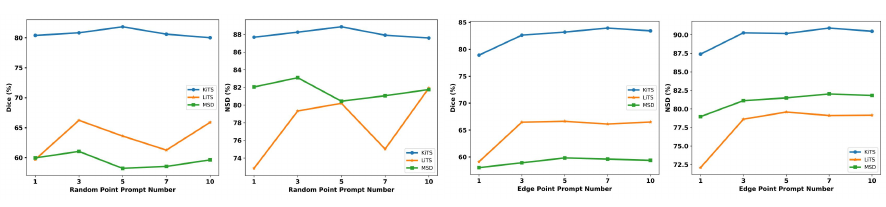}}
\caption{Performance differences between ``random selection" and ``edge selection" as the number of point prompts increase.}
\label{supfig4}
\end{figure*}


\end{document}